\documentclass[iop,usenatbib]{emulateapj}

\usepackage[english]{babel}
\usepackage{graphicx}
\usepackage{fleqn}
\usepackage{amssymb}
\usepackage{amsmath}
\usepackage{booktabs}
\usepackage{hyperref}

\renewcommand{\S}{Section}
\newcommand{\F}{Fig.}

\newcommand{\g}{\mathrm{g}}
\newcommand{\p}{\mathrm{p}}
\newcommand{\m}{\mathrm{m}}

\newcommand{\mj}{M_\mathrm{J}}

\newcommand{\me}{M_\Earth}
\newcommand{\re}{R_\Earth}

\newcommand{\msun}{\mathrm{M}_\odot}

\newcommand{\au}{\textsc{au}}

\newcommand{\capt}{\mathrm{cap}}
\newcommand{\tid}{\mathrm{tides}}
\newcommand{\per}{\mathrm{per}}
\newcommand{\cir}{\mathrm{circ}}
\newcommand{\cur}{\mathrm{cur}}
\newcommand{\syn}{\mathrm{syn}}

\def\apgt{\ {\raise-.5ex\hbox{$\buildrel>\over\sim$}}\ }
\def\aplt{\ {\raise-.5ex\hbox{$\buildrel<\over\sim$}}\ }

\begin{document}

\title{Catching a planet: A tidal capture origin for the exomoon candidate Kepler 1625b I}

\author{Adrian S. Hamers}
\affil{Institute for Advanced Study, School of Natural Sciences, Einstein Drive, Princeton, NJ 08540, USA}
\email{hamers@ias.edu}

\author{Simon F. Portegies Zwart}
\affil{Leiden Observatory, Leiden University, PO Box 9513, NL-2300 RA Leiden, the Netherlands}
\email{spz@strw.leidenuniv.nl}

\begin{abstract}
  The (yet-to-be confirmed) discovery of a Neptune-sized moon around the $\sim 3.2$ Jupiter-mass
  planet in Kepler 1625 puts interesting constraints on the formation
  of the system.  In particular, the relatively wide orbit of the moon
  around the planet, at $\sim40$ planetary radii, is hard to reconcile
  with planet formation theories. We demonstrate that the observed
  characteristics of the system can be explained from the tidal
  capture of a secondary planet in the young system. After a quick phase of tidal circularization,
  the lunar orbit, initially much tighter than $40$ planetary radii, subsequently gradually widened due to tidal synchronization of the spin of the planet with the orbit,
  resulting in a synchronous planet-moon system.  Interestingly, in our scenario the
  captured object was originally a Neptune-like planet, turned into a moon by
  its capture.
\end{abstract}


\section{Introduction}
\label{sect:introduction}

First of its kind discoveries generally put interesting constraints on
our understanding. The first planet \citep{1992Natur.355..145W} as well
as the first Solar system-passing interstellar asteroidal-object
\citep{2017MPEC....U..181B,2017MPEC....U..183M,2017MPEC....W...52M}
surprised many theorists and started a flurry of speculations on their
origin.  A first moon discovered outside the Solar system would also pose
a number of interesting constraints and possibilities for its origin.

A candidate for such a moon (a natural satellite that orbits an exoplanet) was recently found
around the $\sim 1.079\, \msun$ mass star
2MASS~J19414304+3953115 \citep{2017ApJS..229...30M}. Since the
discovery of a $\sim 3.2 \,\mj$ planet in a circular $\sim 0.84\,\au$
orbit, this system is better known as Kepler~1625 (see
\url{https://exoplanets.nasa.gov/newworldsatlas/2271/kepler-1625b/}).

Compelling evidence for a Neptune-like moon orbiting the $\sim 3.2\, \mj$ planet Kepler 1625 b
at a separation of $\sim 40$ planetary radii 
was recently found (\citealt{2018AJ....155...36T,2018SciAdv4.10.1784T}; however, there exists the possibility that the exomoon signal is a false positive, see \citealt{2018A&A...617A..49R}). This hypothetical moon, Kepler 1625b I,
is remarkably massive (with a mass of about 1/100 of the planetary mass) and large compared
to Kepler 1625b, and poses an intriguing problem regarding its
formation. \citet{2018SciAdv4.10.1784T} speculate that its origin challenges
theorists (which is emphasized in \citealt{2018A&A...610A..39H}, who consider a tidal capture scenario, although through planet-binary encounters).

In this Letter, we argue that, although the (hypothetical) moon puts interesting constraints on the
early dynamical evolution of the planet-moon system, its existence
is not surprising.  According to our understanding, the current moon
was born a planet in orbit around the star 2MASS~J19414304+3953115.
This planet turned into a moon upon its tidal capture with the 
more massive planet. Further tidal interaction circularized and widened 
the orbit due to angular-momentum transfer from the spin of the planet to the orbit until synchronization.
For convenience, we will keep referring to ``planet'' for the giant planet Kepler 1625b, and ``moon'' for its
companion Kepler 1625b I, although both should be called planet according to this
scenario.

We demonstrate that this process is feasible, and leads to massive
moons in relatively wide ($\gtrsim 10\,R_{\rm planet}$) orbits around
relatively old ($\gtrsim 1$\,Gyr) stars. In our scenario, we predict that the planet
and moon are currently synchronized with their orbit, and we can put constraints on the 
primordial spin of the planet. 

In \S\,\ref{sect:an}, we consider simple analytic arguments for the conditions of capture, and
investigate the primordial spin of the planet necessary to explain the current orbit. We give an explicit
numerical example of the secular tidal evolution after capture in \S\,\ref{sect:num}. We discuss the likelihood of our scenario
in \S\,\ref{sect:discussion}, and conclude in \S\,\ref{sect:conclusions}.

\section{Analytic estimates}
\label{sect:an}
\begin{figure}
\centering
\includegraphics[scale = 0.6, trim = 0mm 0mm 0mm 0mm]{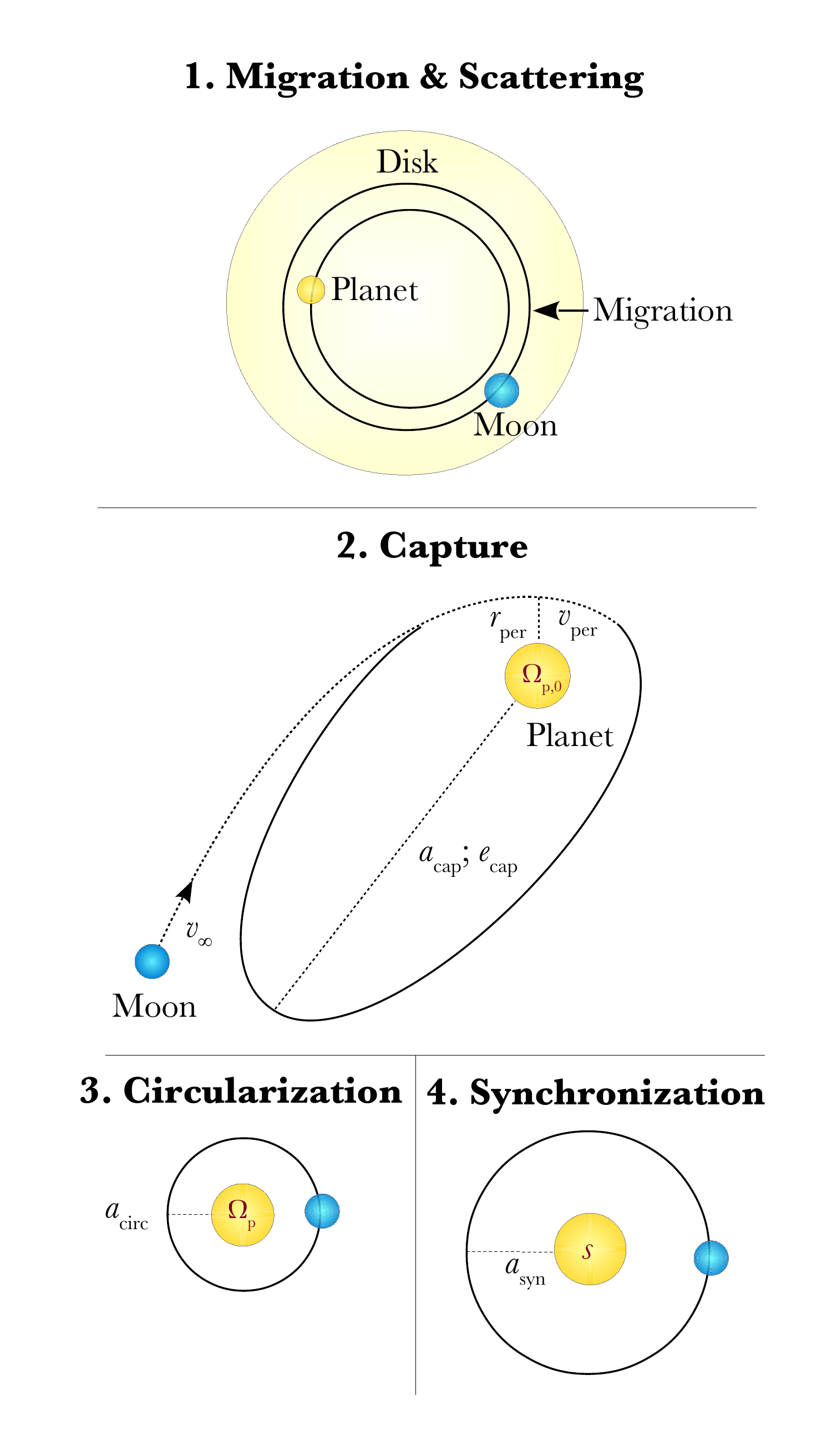}
\caption{\small Sketch of the scenario of tidal capture. In stage 1, the star is not shown, and only one possibility of
convergent migration is shown (the moon outside, and migrating inward). The symbols
  used are described in the text.}
\label{fig:sketch}
\end{figure}

We recognize four distinct stages, which we illustrate in
\F\,\ref{fig:sketch}.

\begin{enumerate}
\item Migration and scattering: two planets embedded in a protoplanetary disk
  migrate towards similar orbits, triggering a short-lived phase of
  dynamical instability.
\item Capture: during the dynamical instability phase, the lighter
  planet (henceforth ``moon'', with mass and radius $M_\m$ and $R_\m$,
  respectively) approaches the more massive planet (with mass and
  radius $M_\p$ and $R_\p$, respectively) to a distance $r_\per$,
  leading to a strong tidal encounter that initiates its capture.
\item Circularization: the moon is initially captured onto a wide and
  highly eccentric orbit (but still within the planet's Hill radius,
  $r_\mathrm{H}$). Tidal dissipation subsequently leads to the circularization
  of the orbit.
\item Synchronization: residual spin angular momentum of the planet
  (spin frequency $\Omega_{\p}$) is gradually transferred to the orbit of the moon
  around the planet, resulting in expansion until synchronization is
  reached\footnote{Since the moment of inertia of the moon is much smaller than that of the
  planet (see below), the moon cannot transfer a significant amount of angular momentum, and
   is quickly synchronized with the orbit.}.
\end{enumerate}

We write the moment of inertia of the planet and the moon as $I_\p =
r_{\g,\p} M_\p R_\p^2$, and $I_\m = r_{\g,\m} M_\m R_\m^2$,
respectively.  Here, $r_{\g,\p}$ is the gyration radius of the planet,
and $r_{\g,\m}$ for the moon, both we assume to have a value of 0.25.
We adopt the ``canonical'' values of $M_\p=10^3\,\me \simeq 3.15\,\mj$, $R_\p=11.4\,\re$
for the planet, and $M_\m=10\,\me \simeq 0.0135\,\mj$, $R_\m=4.0\,\re$ for the moon \citep{2018SciAdv4.10.1784T}. For
these values, $I_\p/I_\m \simeq 812$, and we can safely neglect the
spin angular momentum of the moon.  We furthermore define the reduced mass
$\mu \equiv M_\p M_\m/M$, where $M\equiv M_\p + M_\m$.

\subsection{Conditions for tidal capture}
\label{sect:an:cond}
We assume that the moon approaches the planet on a hyperbolic orbit
with periapsis distance $r_\per$. When the interaction results from
the gradual migration of the planet or moon, both orbits are similar
upon the tidal encounter, and we expect their relative velocity (i.e.,
the hyperbolic velocity at infinity), $v_\infty$, to be small. We set
$v_\infty$ to be a fraction $\alpha$ of the circular orbital
velocity at the separation of the planet+moon system, i.e.,
\begin{align}
\label{eq:v_infty}
v_\infty = \alpha \sqrt{GM_\star/a_\star},
\end{align}
where $M_\star=1.079\,\msun$ is the stellar mass, and $a_\star=0.84\,\au$ \citep{2017ApJS..229...30M}.

The initial orbital energy is $\mu v_\infty^2/2$. For tidal capture to
be successful, sufficient energy should be dissipated in the planet
and moon during the first passage to produce a bound orbit. In
addition, after first passage the apoapsis distance should remain well
within the Hill radius, $r_\mathrm{H}$; otherwise, the star will perturb the newly
captured moon's orbit, preventing its return to the
planet. Approximately, this condition is described by
\begin{align}
\label{eq:cap_cond}
a_\capt < r_\mathrm{H}/2,
\end{align}
were $a_\capt$ is the semimajor axis of the planet-moon orbit directly
after tidal capture. Here,
\begin{align}
r_\mathrm{H} = a_\star \left (\frac{M}{3M_\star} \right )^{1/3},
\end{align}
is the planet's Hill radius. The factor 2 in
equation~(\ref{eq:cap_cond}) takes into account that the captured
orbit is initially highly eccentric; therefore, $r_\mathrm{H}$ should
be compared to the apoapsis distance $a_\capt (1+e_\capt) \approx 2\,
a_\capt$.

We calculate $a_\capt$ from the conservation of energy. Specifically,
consider the initial energy, and the energy after first passage. The
latter consists of the (negative) orbital energy, and the amount of
energy dissipated in the tides, $\Delta E_\tid$ ($\Delta E_\tid>0$). Therefore,
\begin{align}
\label{eq:a_capt}
\frac{1}{2} \mu v_\infty^2 = - \frac{G\mu M}{2a_\capt} + \Delta E_\tid.
\end{align}
We use the formalism of \citet{1977ApJ...213..183P} to compute $\Delta
E_\tid$ in both the planet and moon as a function of the masses,
radii, and the periapsis distance $r_\per$. Specifically, $\Delta E_\tid = \Delta
E_{\tid,\p} + \Delta E_{\tid,\m}$, where
\begin{align}
\label{eq:E_tid_i}
\Delta E_{\tid,i} = \frac{GM_{3-i}^2}{R_i} \sum_{l=2}^3 \left ( \frac{R_i}{r_\per} \right )^{2l+2} T_l(\eta_i),
\end{align}
with 
\begin{align}
\eta_i \equiv \left ( \frac{M_i}{M} \right )^{1/2} \left ( \frac{r_\per}{R_i} \right )^{3/2}.
\end{align}
Here, $M_{3-i}$ is the companion mass. The dimensionless functions
$T_l(\eta_i)$ depend on the structure of the planet/moon. We assume
polytropic pressure-density relations, and adopt analytic fits to
$T_l(\eta_i)$ for polytropic indices of $n=1.5$ or 2 as determined
by \citet{1993A&A...280..174P}. In equation~(\ref{eq:E_tid_i}),
we take the two lowest-order harmonic modes ($l=2$ and $l=3$), which give a good
description \citep{1977ApJ...213..183P}. 

The analytic fits for $T_l(\eta)$ from \citet{1993A&A...280..174P} do not account for the planetary and lunar spins. In the case of significant spins, however, $T_l(\eta)$ could be a few times larger. For simplicity, we ignore this complexity, but note that this adds some uncertainty to our calculation of $a_\capt$. 

\begin{figure}
\centering
\includegraphics[scale = 0.48, trim = 5mm 0mm 0mm 0mm]{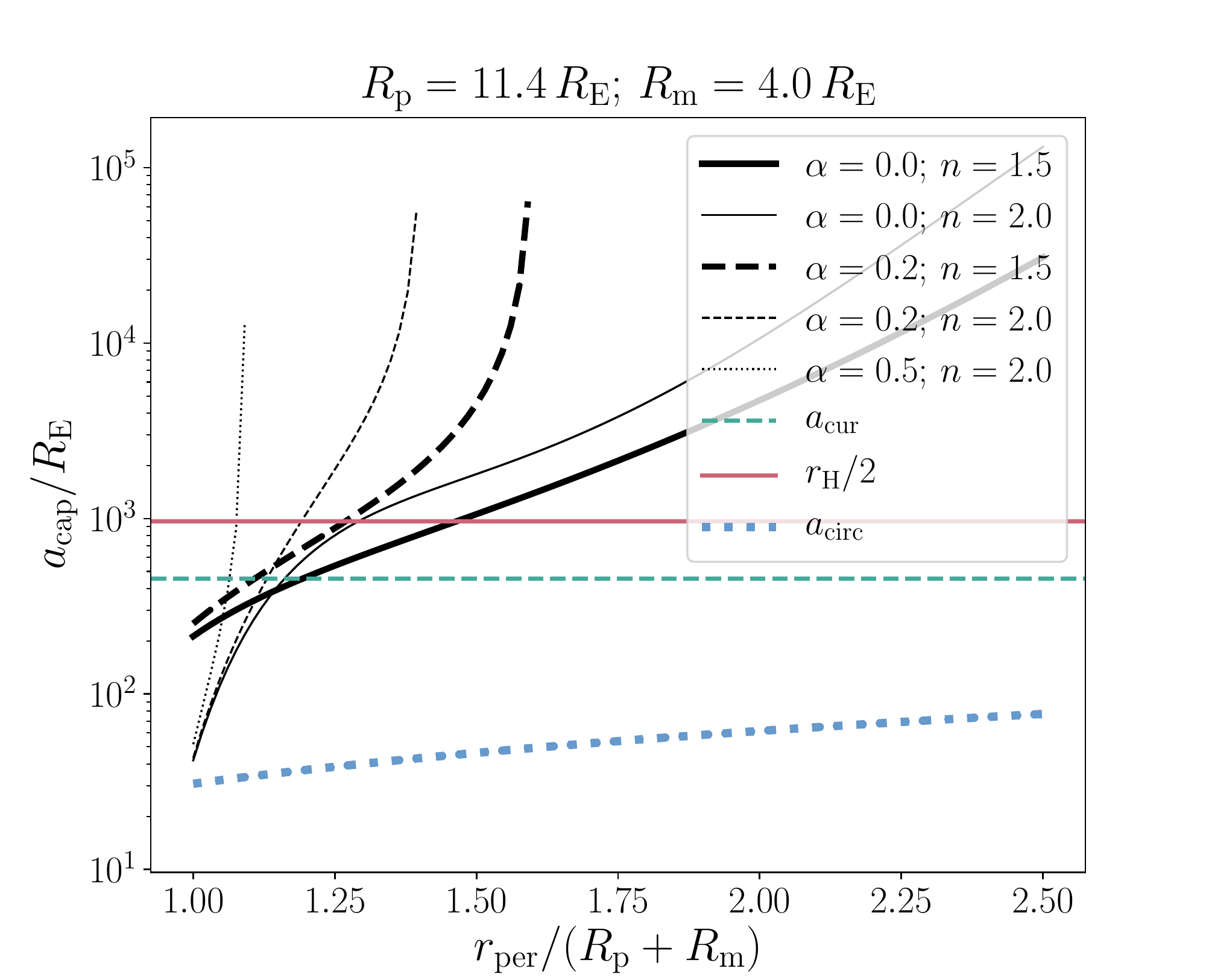}
\caption{\small Capture semimajor axis $a_\capt$ as a function of
the periapsis distance $r_\per$ according to equation~(\ref{eq:a_capt}). The canonical
  radii are assumed, with different combinations of $v_\infty$
  (quantified by $\alpha$) and the polytropic index $n$. The red solid (green dashed)
  horizontal line shows $r_\mathrm{H}/2$ ($a_\mathrm{cur}$, the current semimajor axis). The blue
  dotted line shows $a_\cir$ (see equation~\ref{eq:a_circ}). }
\label{fig:a_cap}
\end{figure}

In \F\,\ref{fig:a_cap}, we plot $a_\capt$ as a function of $r_\per$
according to equation~(\ref{eq:a_capt}). We assume the canonical
radii, and consider different combinations of $v_\infty$ (quantified
by $\alpha$), and the polytropic index $n$ (a larger $n$ corresponds
to a more centrally-concentrated planet/moon). A polytropic index of
$n=1.5$ is a reasonable approximation for the structure of a gas giant
planet \citep{2015MNRAS.452.1375W}. The red solid (green dashed)
  horizontal line shows $r_\mathrm{H}/2$ ($a_\mathrm{cur}$, the current semimajor axis,
  which we set to $a_\mathrm{cur} = 40\,R_\p = 456\,\re$). With our parameters, tidal energy dissipation during the capture is dominated by the moon, with $\Delta E_{\tid,\m}/\Delta E_\tid \simeq 0.94$ for $r_\per/(R_\p+R_\m)=1$, and increasing to $\Delta E_{\tid,\m}/\Delta E_\tid \simeq 0.98$ for $r_\per/(R_\p+R_\m)=1.5$. 
  
\begin{figure}
\centering
\includegraphics[scale = 0.48, trim = 5mm 0mm 0mm 0mm]{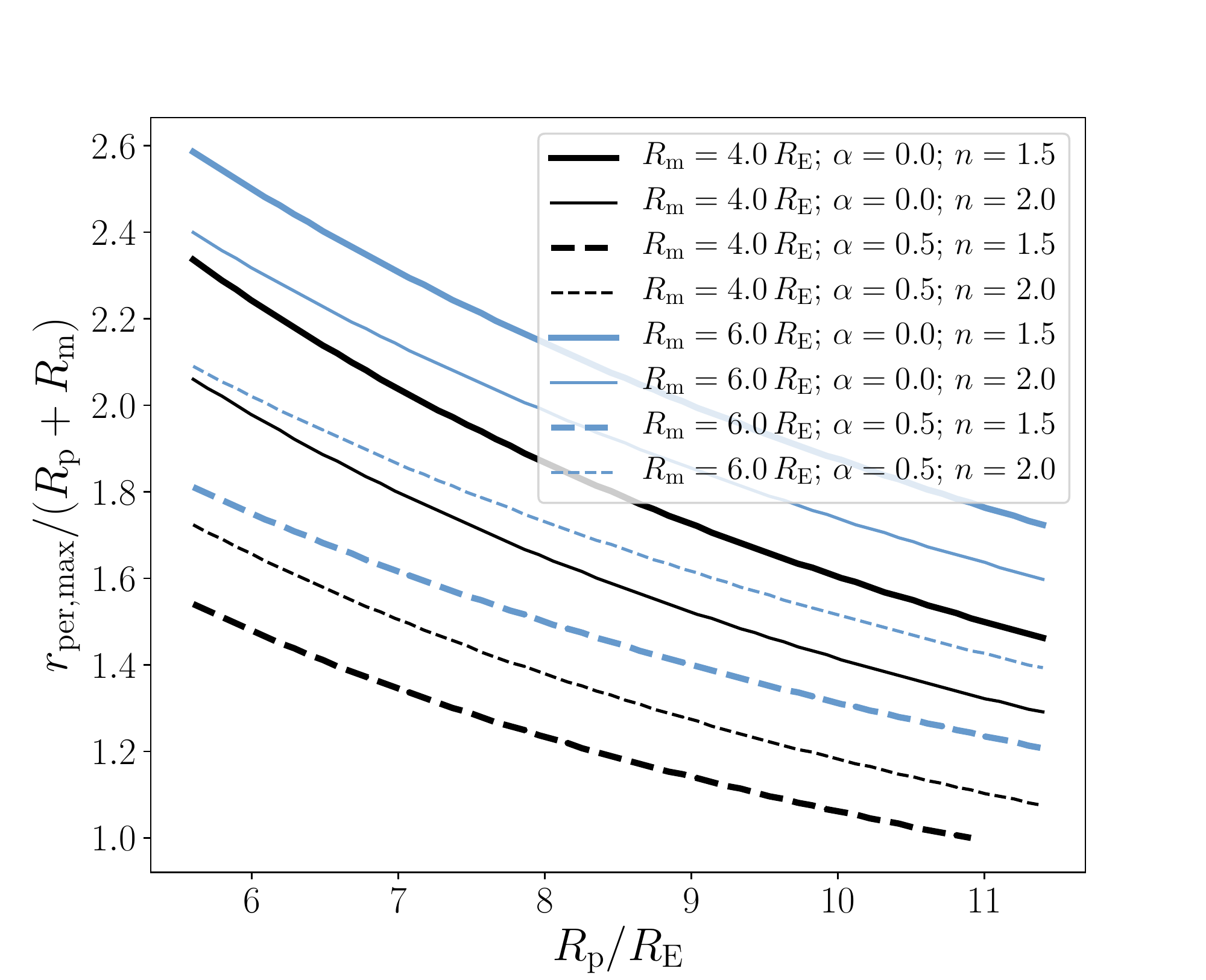}
\caption{\small Largest periapsis distance for which
capture is possible, $r_\mathrm{per,max}$, plotted as a function of $R_\p$, and for different combinations
of $R_\m$, $\alpha$, and $n$. }
\label{fig:rp_max}
\end{figure}

For sufficiently small $r_\per$, the moon can be tidally captured
without its orbit being perturbed by the star. The range in $r_\per$
is typically small, but increases for smaller $v_\infty$ (i.e., smaller $\alpha$) and smaller
$n$. The range of $r_\per$ increases for a smaller planet.  This is
shown explicitly in \F\,\ref{fig:rp_max}, in which the largest
periapsis distance for which capture is possible,
$r_\mathrm{per,max}$, is plotted as a function of $R_\p$, for
different combinations of $R_\m$, $\alpha$, and $n$.

\subsection{Orbital expansion due to secular tidal evolution}
\label{sect:an:exp}
After tidal capture, the orbit is highly eccentric. Subsequently, the orbit orbit shrinks and circularizes. 
The semimajor axis after circularization can be estimated as
\begin{align}
\label{eq:a_circ}
a_\cir \simeq 2 \, r_\per.
\end{align}
Tidal capture alone cannot explain the current orbit of the
planet-moon system in Kepler 1625. This is exemplified in
\F\,\ref{fig:a_cap}, in which $a_\cir$ is shown with the blue dotted
curves. For any reasonable values of $r_\per$, $a_\cir$ is smaller
then the currently observed semimajor axis, $a_\cur$, by about an
order of magnitude. Here, we set $a_\cur$ to $40\,R_\p =
456\,\re$.

After capture, the expansion of the orbit to the currently observed
orbit is mediated by the transfer of angular momentum from the spin of
the planet to the orbit. This process continues until the planet and
orbit are in synchronous rotation (analogous to the current tidal
evolution of the Earth-Moon system).

Using the fact that angular momentum is conserved during the entire process (capture, circularization, and synchronization),
we can equate the initial angular momentum before capture to the angular momentum after synchronization. After synchronization, the planetary spin-frequency is equal to
the orbital frequency, $s=\sqrt{GM/a_\syn^3}$, where $a_\syn$ is the semimajor axis of the synchronized orbit. Therefore, neglecting the moon's spin angular momentum,
\begin{align}
\label{eq:AM_cap}
\mu v_\per r_\per + I_\p \Omega_{\p,0} = \mu \sqrt{GMa_\syn} + I_\p \sqrt{\frac{GM}{a_\syn^3}}.
\end{align}
Here, $\Omega_{\p,0}$ is the spin frequency of the planet before the
tidal encounter (i.e., the primordial spin frequency), and $v_\per$ is the orbital speed at periapsis at first
approach. We compute $v_\per$ by assuming a purely hyperbolic orbit on first approach, i.e.,
\begin{align}
\label{eq:v_per}
v_\per = \sqrt{v_\infty^2 + \frac{2GM}{r_\per}}.
\end{align}
Writing the initial planet's spin as $\Omega_{\p,0} = \beta \sqrt{GM_\p/R_\p^3}$, where $\beta$ is a dimensionless parameter that measures the initial
planetary spin in units of its breakup rotation rate, we obtain from equations~(\ref{eq:v_infty}), (\ref{eq:AM_cap}), and (\ref{eq:v_per}) 
the following expression for the minimum required spin of the planet such that the synchronized orbit has semimajor axis $a_\syn$,
\begin{align}
\label{eq:beta}
\nonumber \beta &= \sqrt{\frac{M}{M_\p}} \left [ \left ( \frac{R_\p}{a_\syn} \right )^{3/2} + \frac{M_\m}{M} r^{-1}_{\g,\p} \left ( \sqrt{\frac{a_\syn}{R_\p}} \right. \right. \\
&\quad \left. \left. - \sqrt{\frac{2r_\per}{R_\p}} \sqrt{1 + \alpha^2 \frac{M_\star}{M} \frac{r_\per}{2 a_\star} } \right ) \right ].
\end{align} 

After circularization, the orbit asymptotically evolves to
synchronization, expanding the orbit in the process. The associated
timescale depends on the efficiency of tidal dissipation (see \S\,\ref{sect:num} below). We expect
the currently observed orbit to be close to synchronization.
Therefore, by setting $a_\syn = a_\cur$, we can use
equation~(\ref{eq:beta}) to determine, as a function of $r_\per$, the
minimal initial planetary spin (quantified by $\beta$) required to
explain the currently observed orbit.

\begin{figure}
\centering
\includegraphics[scale = 0.45, trim = 0mm 0mm 0mm 0mm]{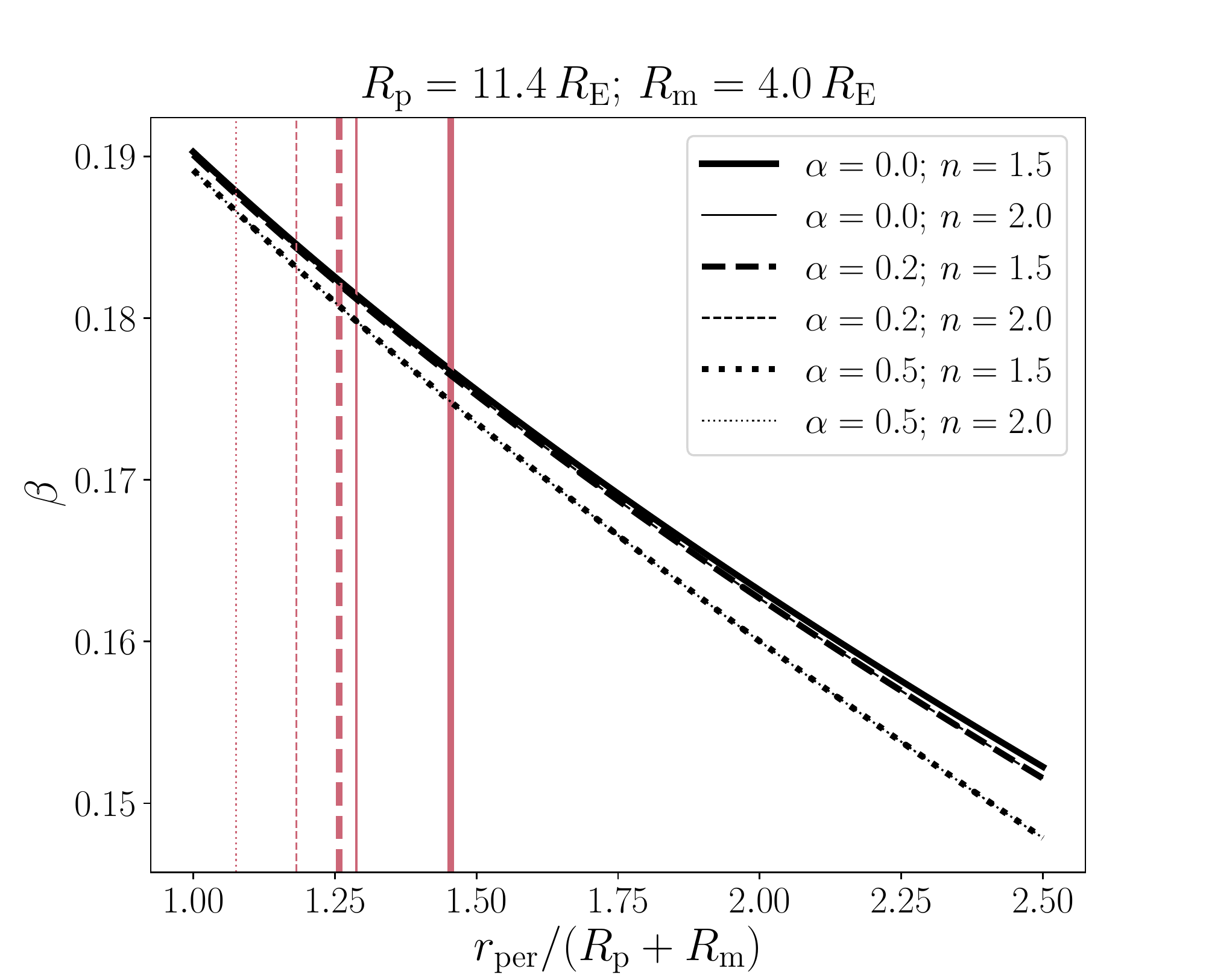}
\caption{\small Initial planetary spin (quantified by the fraction $\beta$ of breakup rotation) required to explain the current orbit of Kepler 1625 through tidal capture, as a function of $r_\per$. Assumed radii are $R_\p=11.4\,\re$, and $R_\m=4\,\re$. Different line styles and thicknesses correspond to different $\alpha$ and $n$. The vertical red lines indicate the maximum $r_\per$ for which capture can be successful. }
\label{fig:beta}
\end{figure}

In \F\,\ref{fig:beta}, we present the resulting values for $\beta$ for
a selection of values for $\alpha$ and $n$.  The vertical lines (in
red) indicate the maximum value of $r_\per$ below which capture can be
successful, i.e., $a_\capt<r_\mathrm{H}/2$, assuming $R_\p=11.4\,\re$,
and $R_\m=4\,\re$. The dependence of $\beta$ on $r_\per$ is not
strong; generally, $\beta\sim0.2$, i.e., $20\%$ of breakup rotation is
required. For $R_\p=5.6\,\re$ and $R_\m=4\,\re$ (not shown here), the allowed (normalized) range in
$r_\per/(R_\p+R_\m)$ is larger, but the required $\beta$ to explain the current orbit
is larger; typically, $\beta\sim 0.3$. The minimum value for
$\beta$ is lower for non-zero $\alpha$ (in which case some angular
momentum can be transferred from the initial orbit to the planetary
spin), but the differences between $\alpha=0$ and $\alpha=0.5$ are
small.

A rotation rate of a few tens of per cent of breakup rotation is not extreme nor unusual. For example, Jupiter, Saturn and Neptune are rotating at $\simeq 0.3$, $0.4$ and $0.2$ of breakup rotation, respectively. Massive Jupiter-like extrasolar planets are known to have similar rotation rates (see, e.g., Fig. 2 of \citealt{2018NatAs...2..138B}). 

\section{Numerical example of secular tidal evolution}
\label{sect:num}
In \S\,\ref{sect:an}, we derived analytic expressions for the tidal
evolution of the planet-moon system after capture. Here, we illustrate
the long-term tidal evolution that results from the capture of the
moon by the planet by integrating the secular equations of motion
numerically.

We adopt the equilibrium tide model by \cite{1998ApJ...499..853E},
with the apsidal motion constants
$k_\mathrm{AM,p}=k_\mathrm{AM,m}=0.19$. For the tidal time-lags, we
adopt either $\tau=\tau_\p=\tau_\m = 6.6\,\mathrm{s}$ or 66 s. A value of $\tau = 6.6\,\mathrm{s}$ corresponds to 10 times longer
(i.e., stronger tides) than $0.6\,\mathrm{s}$, as inferred to be
appropriate for high-eccentricity migration by
\citet{2012arXiv1209.5724S}. These efficient tides turn out to be necessary to explain the current orbit with our nominal parameter values (see below). For simplicity, we use the equilibrium tide model
for the evolution immediately after capture, when the eccentricity is still high ($e>0.9$). A caveat of this is that the equilibrium tide model does not accurately describe the evolution for eccentricities $\gtrsim 0.8$ \citep{1995ApJ...450..732M}.  

\begin{figure}
\centering
\includegraphics[scale = 0.32, trim = 18mm 10mm 0mm 0mm]{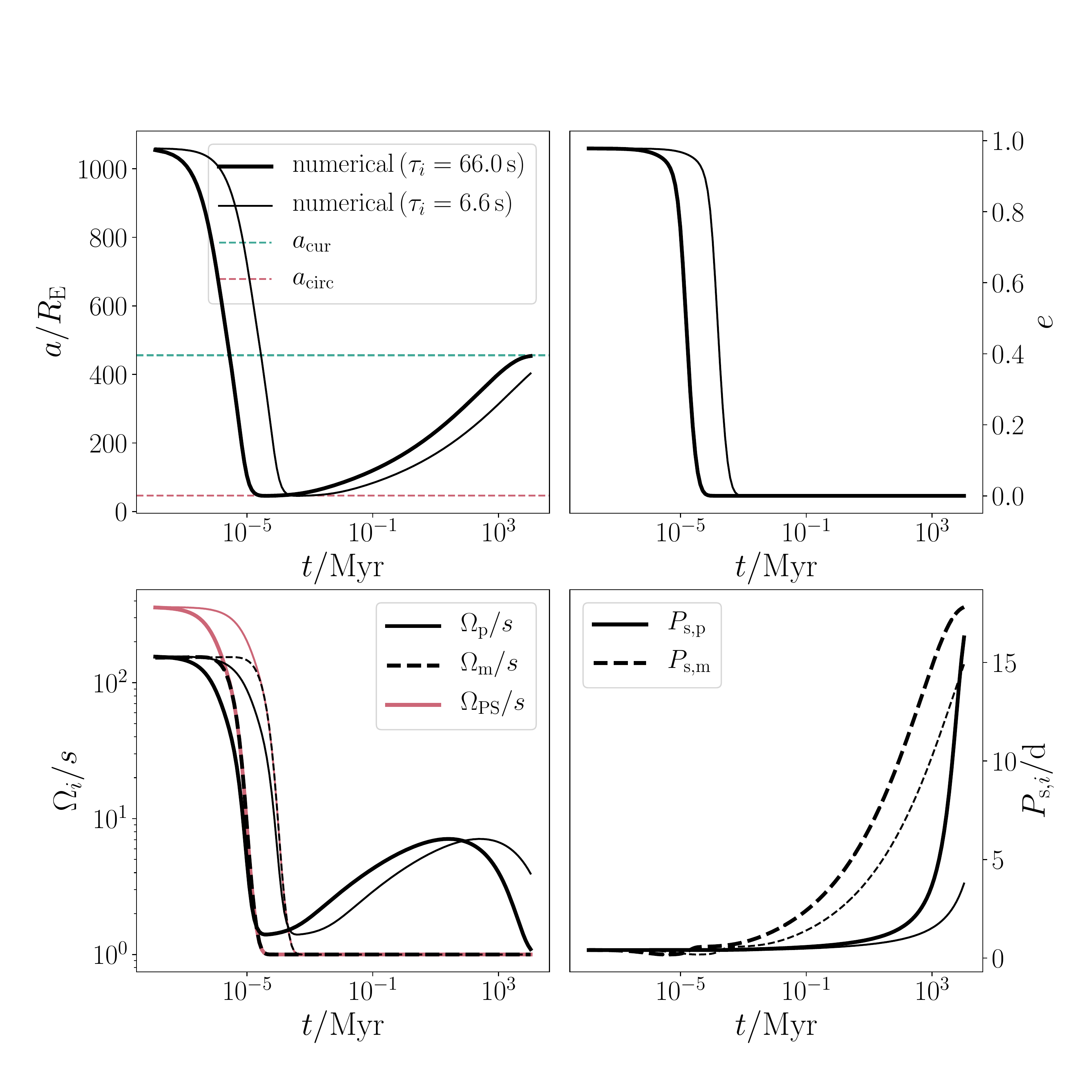}
\caption{\small Long-term evolution of the semimajor axis (top-left panel), eccentricity (top-right panel), the spin rates normalized to the orbital mean motion $s$ (bottom-left panel), and the rotation periods (bottom-right panel) in a tidal capture scenario for Kepler 1625, obtained by numerically integrating the secular tidal equations of motion. Thick and thin lines correspond to a time lag of 66 and 6.6 s, respectively. In the top-left panel, the green dashed line indicates the current semimajor axis of the planet-moon orbit; the red dashed line shows the expected circularization semimajor axis, equation~(\ref{eq:a_circ}). In the bottom-left panel, the red lines show the expected curves for pseudosynchronous rotation computed using equation 42 of \citet{1981A&A....99..126H} and the eccentricity as a function of time from the numerical simulations. }
\label{fig:secular_tides_num}
\end{figure}

We start the integration with a semimajor axis of $a_0\simeq
1060\,\re$.  This value corresponds to (borderline) capture at
$r_\per=1.5\,(R_\p+R_\m)$ with $\alpha=0$ and $n=1.5$ (see
\F\,\ref{fig:a_cap}). The corresponding initial eccentricity is $e_0 =
1-r_\per/a_0 \simeq 0.978$. According to our analytic estimates (see
\F\,\ref{fig:beta}), the critical planetary spin for the final orbit
to match the current orbit is $\beta \simeq 0.1756$. We adopt this
spin rate for the planet. The rotation period of the moon is set to
$10\,\mathrm{hr}$; note, however, that the latter does not affect the
synchronized semimajor axis since $I_\p \gg I_\m$. In the numerical
integrations, the spins are assumed to be initially aligned with the
orbit.

We present in \F\,\ref{fig:secular_tides_num}, the time evolution of
the semimajor axis, eccentricity, and the spin rates, where the integration lasts for 10 Gyr, approximately the age of the star \citep{2018SciAdv4.10.1784T}. The thick and thin lines correspond to
a time lag of 66 and 6.6 s, respectively. The initial
evolution is rapid, circularizing and shrinking the orbit to a value
which is consistent with $a_\cir$ (red dashed line in the top-left panel; see equation~\ref{eq:a_circ})
within $\sim 10\,\mathrm{yr}$.  The moon, which has a small moment of
inertia, is synchronized within the same time span, whereas the planet
remains spinning more rapidly than the orbit for up to $\sim
10\,\mathrm{Gyr}$. The planet is synchronized within $\sim 10\,\mathrm{Gyr}$ assuming extremely
efficient tides ($\tau=66\,\mathrm{s}$), and the steady-state semimajor axis is consistent with the currently observed value (green dashed line; this is consistent with the analytic expression for $\beta$ presented in
equation~\ref{eq:beta}). Assuming less efficient tides ($\tau=6.6\,\mathrm{s}$), equilibrium is not yet reached after 10 Gyr, although it is close ($a$ reaching $a_\cur$ within $\simeq 13\%$). Evidently, even weaker tides would make the agreement with the current orbit within 10 Gyr more difficult. 

We can estimate the time-scales for circularization and synchronization analytically as follows. Circularization is dominated by the tides in the moon, and during the circularization phase, the spins are quickly brought to pseudosynchronous rotation (see equation 42 of \citealt{1981A&A....99..126H}, and the red lines showing $\Omega_\mathrm{PS}/s$ in the bottom-left panel of \F\,\ref{fig:secular_tides_num}). Also taking the limit $e\rightarrow 1$, the circularization time-scale can then be estimated as \citep{1981A&A....99..126H}
\begin{align}
\label{eq:t_e}
\nonumber t_e &\equiv \left ( \frac{1}{e} \frac{\mathrm{d}e}{\mathrm{d} t} \right )^{-1} \\
\nonumber &\sim (1-e^2)^{13/2} \left (\frac{a_0}{R_\m} \right )^8 \frac{T_\m}{k_\mathrm{AM,\m}} \frac{1}{q_\m(1+q_\m)} \frac{1}{27} \frac{320}{451} \\
&\simeq 9 \times 10^2 \,\mathrm{yr}.
\end{align}
Here, $T_i \equiv R^3_i/(GM_i\tau_i)$ \citep[equation 12]{1981A&A....99..126H}, and $q_i = M_{3-i}/M_i$. In the last line of equation~(\ref{eq:t_e}), we substituted numerical values, assuming $\tau_i=66\,\mathrm{s}$. The resulting time-scale is roughly consistent with the circularization time-scale in the numerical example.  

To estimate the synchronization time-scale, we take advantage of the separation of time-scales for circularization and synchronization. After circularization, $e=0$ and $a=a_\cir\simeq 2 \,r_\per=46.2\,\re$, whereas the planetary spin (which dominates the spin angular-momentum budget) is still equal to its initial value to good approximation (see the bottom-right panel of \F\,\ref{fig:secular_tides_num}). In this case, one can show using equations (9) and (11) of \citet{1981A&A....99..126H} that $a$ and $\Omega_\p$ are related according to
\begin{align}
a^{1/2}-a_\cir^{1/2} = - C(\Omega_\p-\Omega_{\p,0}),
\end{align}
where $C \equiv (1+q_\p) r_{\mathrm{g},\p} R_\p^2/(q_\p \sqrt{GM})$. 
By integrating the equation for $\mathrm{d}a/\mathrm{d} t$ over time, we find a synchronization time-scale
\begin{align}
\nonumber t_\Omega &\equiv \int_{a_\cir}^{a_\mathrm{f}} \frac{\mathrm{d} a}{\dot{a}} = \frac{1}{6} \frac{T_\p}{k_\mathrm{AM,\p}} \left ( \frac{a_\cir}{R_\p} \right )^8 \frac{1}{q_\p(1+q_\p)} \\
\nonumber &\quad \times \int_1^{a_\mathrm{f}/a_\cir} \frac{x^7\,\mathrm{d} x}{x^{3/2} \left [A-B\left(x^{1/2}-1\right ) \right ]-1} \\
&\simeq 2\times 10^{10}\,\mathrm{yr}.
\end{align}
Here, $A \equiv \Omega_{\p,0} \sqrt{a_\cir^3/(GM)}$, $B \equiv a_\cir^2/(C \sqrt{GM})$, and $a_\mathrm{f}$ is the final semimajor axis. For the numerical estimate, we again assumed $\tau_i=66\,\mathrm{s}$, whereas we set $a_\mathrm{f} = 0.99 \, a_\cur$. The numerical value is roughly consistent with the synchronization time-scale in \F\,\ref{fig:secular_tides_num}.

\section{Discussion}
\label{sect:discussion}
Multiple bodies form during the early evolution of a debris disk to a
fully populated planetary system. The migration of planets in such an
environment is to be expected, in particular when the residual gas
causes a drag force on the planets. The efficiency of this drag force
is proportional to the planet mass \citep{2015A&A...574A..52D}. For
a tidal capture to become possible, the two planets have to acquire
similar orbits, which can be realized via drag. It remains unclear if
the more massive planet was born further out and migrated inwards to
the lower-mass planet, or that the more massive planet originally
orbited closer to the star. In the latter case, the disk must have
had an inner edge to prevent the inner more massive planet to migrate
further inwards. 

In both cases, the encounter is expected to occur
with comparable orbits, i.e., the encounter is parabolic, or hyperbolic with a 
relatively low speed at infinity. The outcome
of this encounter can be the ejection of one of the planets (most likely the lower-mass
planet), collisions with the star, tidal capture, or a collision of the planets with each other.
We can estimate the branching ratios between these scenarios by comparing the
relevant cross-sections. Here, we do not consider collisions with the star.

For ejections to occur, we require the velocity change imparted on the lower-mass planet 
(mass $M_\m$) during the encounter at a distance of $\sim a_\star$ to be comparable to the local escape velocity from the star, i.e., 
$\Delta v_\m \sim v_\mathrm{esc} = \sqrt{2GM_\star/a_\star}$. The (3D) velocity change for an encounter with impact
parameter $b$ can be estimated as (e.g., \citealt[S3.1(d)]{2008gady.book.....B})
\begin{align}
\Delta v_\m \approx \frac{2M_\p}{M} \frac{v_\infty}{\sqrt{1+\left (b/b_{90} \right )^2}},
\end{align}
where $b_{90} \equiv GM/v_\infty^2 = (a_\star/\alpha^2) (M/M_\star) \simeq (55.4/\alpha^2) \,\re$ is the
impact parameter for a $90^\circ$ deflection. For $\alpha\sim 1$, $b_{90} > R_\p+R_\m$, showing that
gravitational focusing is important. The impact parameter for escape can therefore be written as
\begin{align}
b_\mathrm{ej} = b_{90}  \sqrt{2\alpha^2 (M_\p/M)^2 -1}.
\end{align}
Note that $v_\infty$ needs to be large enough for the lower-mass planet to be ejected; specifically, $\alpha \geq \sqrt{(1/2) (M/M_\p)} \simeq 0.71$. 

The impact parameter for tidal capture or direct collision, taking into account gravitational focusing, is
\begin{align}
b = \sqrt{r^2 + \frac{2GMr}{v_\infty^2}} = r \sqrt{1+ 2\frac{b_{90}}{r}},
\end{align}
where we set $r=\gamma (R_\p+R_\m)$ for tidal capture, and $r=R_\p+R_\m$ for direct collision. From 
our analytical estimates (\S\,\ref{sect:an}), $\gamma \lesssim 2.5$ for a successful capture, depending on the parameters (see \F\,\ref{fig:rp_max}). 

Therefore, the branching ratio between capture and ejection is
\begin{align}
\label{eq:branching_ratio_cap_ej}
\nonumber \frac{b_\mathrm{cap}^2}{b_\mathrm{ej}^2}  &= \gamma^2 \frac{(R_\p+R_\m)^2}{b_{90}^2} \frac{1 + 2\frac{b_{90}}{\gamma(R_\p+R_\m)}}{2\alpha^2 (M_\p/M)^2 - 1} \\
\nonumber &\approx \frac{2 \gamma}{2\alpha^2(M_\p/M)^2-1} \frac{R_\p+R_\m}{b_{90}} \\
\nonumber &= \frac{2 \gamma \alpha^2}{2\alpha^2(M_\p/M)^2-1} \frac{M_\star}{M} \frac{R_\p+R_\m}{a_\star} \\
&\simeq \frac{0.56\,\gamma \alpha^2}{2\alpha^2 (M_\p/M)^2-1},
\end{align}
where in the second line we used that $b_{90} > R_\p+R_\m$, and in the fourth line we substituted our adopted values for the masses, the radii, and $a_\star$. For $\alpha=1$ and $\gamma=2.5$, the first line of equation~(\ref{eq:branching_ratio_cap_ej}) gives $b_\mathrm{cap}^2/b_\mathrm{ej}^2 \simeq 1.9$; for $\alpha=0.8$ and $\gamma=2.5$, we get $b_\mathrm{cap}^2/b_\mathrm{ej}^2 \simeq 4.3$.

The branching ratio between capture and collision is
\begin{align}
\label{eq:branching_ratio_cap_col}
\frac{b_\mathrm{cap}^2}{b_\mathrm{col}^2}  = \gamma^2 \frac{1+ \frac{2}{\gamma} \frac{b_{90}}{R_\p+R_\m}}{1+ 2 \frac{b_{90}}{R_\p+R_\m}} \approx \gamma,
\end{align}
where we again used that $b_{90} > R_\p+R_\m$. For $\alpha=1$ and $\gamma=2.5$, the non-approximated equation~(\ref{eq:branching_ratio_cap_col}) gives $b_\mathrm{cap}^2/b_\mathrm{col}^2 \simeq 3.0$ (for $\alpha=0.8$ and $\gamma=2.5$, $b_\mathrm{cap}^2/b_\mathrm{col}^2 \simeq 2.8$). We note that our distinction here between capture and collision is simplistic; e.g., \citet{2018ApJ...852...41H} show using hydrodynamic simulations that interactions with $r_\per/(R_\p+R_\m)<1$ can lead to bound pairs of planets/moons, in addition to mergers. 

We conclude that the likelihoods for ejection, capture, and collision are
comparable within a factor of a few. This is
 consistent with the more detailed calculations of \citet{2010arXiv1007.1418P} and \citet{2014ApJ...790...92O}, who carried out numerical scattering experiments and found roughly equal 
 ejection and capture fractions. The distribution of the
relative inclination of captured planets binary is flat
\citep[see][]{2014ApJ...790...92O}, making the currently observed
$\sim 45^\circ$ angle of the planet-moon orbit with respect to the ecliptic
not unlikely.

We remark that we assumed constant sizes and static interior structure of the planet and moon. If these properties were allowed to vary due to planetary evolution, the synchronization process could occur differently. In particular, the semimajor axis could stall \citep{2017MNRAS.471.3019A}, which would reduce the likelihood that planet-spin-boosted tidal capture can explain the current orbit of Kepler 1625b I. 

Another caveat is that during the migration-induced dynamical instability phase, there could be multiple encounters before a successful capture. During each of these encounters, the system could be disrupted, thereby lowering the capture probability. More detailed $N$-body integrations to take this into account are left for future work.

\section{Conclusions}
\label{sect:conclusions}

We argued that the planet-moon system in Kepler 1625 is the result of
the tidal capture of a secondary planet by the primary planet around
the star. As a result of scattering induced by convergent migration in a disk,
the two planets approached each other on a low-energy hyperbolic or parabolic orbit, and passed each other within 
$\lesssim 2.5\,(R_\p + R_\m)$. The tidal dissipation induced in this encounter subsequently led to the
capture of the minor planet by the primary planet, turning the former
into a moon. The first tidal encounter led to a highly eccentric and
wide orbit, and for capture to be successful, the apocenter should have remained within the planet's Hill sphere. 
The orbit then circularized to a tight orbit, in $\sim 10$ yr. Over a much longer time-scale of $\sim 10$ Gyr, the
primary planet subsequently transferred its spin angular momentum to the orbit, widening the
latter until synchronization. We find that the primary planet must have had a primordial spin of at least $\sim 20\%$ of
critical rotation in order to deposit sufficient angular momentum into the planet-moon
orbit to be consistent with the current orbit. We expect that the current orbit evolves
very slowly, and that both the planet and moon are in almost synchronous
rotation with the orbit. 

These captures are probably not uncommon,
being roughly as common as planet collisions. However, the precise
frequency for this process to operate remains unclear. We
expect that moon formation from tidal capture is not uncommon
\cite[see also][]{2010arXiv1007.1418P,2014ApJ...790...92O}, and probably comparable to the
number of planet collisions or ejections.

The capture must have occurred early in the planetary system's
evolution (more than a Gyr ago) to allow tidal dissipation to
synchronize the system to its current orbit. Our scenario can be
tested by measuring the spins of both planet and moon, which should be
synchronous with the orbit, and along the same axis as the orbital angular momentum of the
planet-moon system.

\section*{Acknowledgements}
We thank Jaime Alvarado-Montes, Ren\'{e} Heller, David Kipping, and Jean Schneider for comments and discussions, and the anonymous referee for a very helpful report. ASH gratefully acknowledges support from the Institute for Advanced Study, and the Martin A. and Helen Chooljian Membership. SPZ thanks Norm Murray and CITA for the hospitality during his long-term visit.

\bibliographystyle{apj}
\bibliography{literature}

\end{document}